\documentclass[a4paper,12pt]{article}

\usepackage[english]{babel}
\usepackage[dvips]{graphicx}
\usepackage{verbatim}
\usepackage{makeidx}
\usepackage{color}
\usepackage{amsfonts}
\usepackage{amsmath}

\newcommand{\GeV}{\mathrm{GeV}}

\newcommand{\mum}{\mathrm{\mu m}}

\title{Detection of $D^0$ mesons {\it via} hadronic decays\\ 
       in Pb--Pb collisions at LHC with ALICE}

\author{A. Dainese \\ {\small for the ALICE Collaboration}\\
{\small Universit\`a degli Studi di Padova, via F. Marzolo 8, 35131 Padova, Italy}\\ 
{\small e-mail: andrea.dainese@pd.infn.it}}

\date{}

\begin{document}

\maketitle

\begin{abstract}
  \noindent
  The ALICE experiment is devoted to the
  study of heavy-ion collisions at the CERN LHC collider.  
  We present the results of a feasibility study for the detection of
  $D^0\to K^-\pi^+$ decays in Pb--Pb collisions with ALICE.
\end{abstract}

\section{Physics motivation}

The aim of the ALICE~\cite{tp} 
experiment is to study the
behaviour of nuclear matter in the conditions of high densities and 
temperatures in which a transition to a deconfined QCD phase 
(Quark Gluon Plasma) is expected. 

The study of open charm production is of primary interest for two 
main reasons:

\begin{itemize}
  \item the interaction of the produced charm quarks with
   the plasma may reduce their momenta because of elastic collisions
   and in-medium gluon emissions (see {\it e.g.} Ref.~\cite{lin} and 
   references therein); 
  \item secondary parton scattering in the high-density
   partonic system produced may provide an additional source of charm
   quarks~\cite{muller}.  
\end{itemize}

The measurement of the total charm production cross section and of 
the transverse momentum distribution of charm quarks in Pb--Pb 
collisions, as well as in pp and in pA interactions, is essential 
to study these open issues.
The exclusive reconstruction of $D^0$ mesons in the hadronic decay channel 
($D^0\to K^-\pi^+$, branching ratio $=3.83\%$) will provide 
a direct measurement of charm kinematical distributions.  

\section{Detection strategy}

\begin{figure}[!t]
  \begin{center}
    \includegraphics[clip,angle=-90.,width=.6\textwidth]{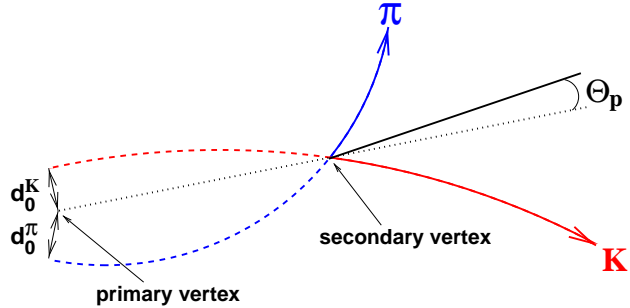}
    \caption{Schematic representation of the $D^0\to K^-\pi^+$ decay with the 
             impact parameters ($d_0$) and the pointing angle ($\theta_P$).} 
    \label{fig:D0sketch}
  \end{center}
\end{figure}

The $D^0$ meson decays through a weak process and has a mean
proper length $c\tau = (123.7\pm 0.8)\ \mum$. 
Therefore, the distance between the interaction point (primary vertex) and 
the decay point (secondary vertex) is measurable. The selection of the 
$D^0\to K^-\pi^+$ decay (and charge conjugate) allows the direct 
identification of the $D^0$ particles 
by computing the invariant mass of fully-reconstructed topologies originating
from displaced secondary vertices. Figure~\ref{fig:D0sketch} shows a sketch 
of the decay: the main feature of this topology is the presence of two tracks
with impact parameters of the order of $100\ \mum$, 
the impact parameter ($d_0$) being the distance 
of closest approach of a particle trajectory to the primary vertex.

The identification of these topologies requires precise measurements
of the tracks momenta and impact parameters. Charged-particle tracks 
are reconstructed using the Time Projection Chamber and the Inner 
Tracking System, which provide measurements of the momentum and of the 
impact parameter of the $D^0$ decay products with resolutions 
$\sigma(p)/p\sim 1\%$ and $\sigma(d_0)\simeq 50\ \mum$, respectively,
using a magnetic field of 0.4 T. Particle identification {\it via} 
time-of-flight allows to reject a large fraction of ($\pi^\pm$,$\pi^\mp$) 
pairs, thus significantly reducing the large combinatorial background of 
opposite-charge track pairs from the underlying high-multiplicity 
Pb--Pb event. 

\section{Analysis and results}

The $c\overline{c}$ production rate in central Pb--Pb collisions at the 
LHC is estimated~\cite{notehvq} from next-to-leading order pQCD~\cite{hvqmnr}
to be 
$N(c\overline{c})=115/$event; this gives $0.53$ $D^0$ mesons per unit of 
rapidity decaying in the $K\pi$ channel.

The charm signal is generated using PYTHIA~\cite{pythia}, while the underlying
\mbox{Pb--Pb} events are generated using HIJING~\cite{hijing}, which gives a 
multiplicity of about $6000$ charged particles per unit of rapidity.

The initial value of the signal-to-background ratio is 
\mbox{$S/B\simeq4.5\cdot 10^{-6}$}. The most effective
selection in order to extract the charm signal out of the large 
combinatorial background is based on the requirement to have two 
tracks with large impact parameters and a good pointing of the reconstructed
$D^0$ momentum to the collision point ({\it i.e.} the pointing angle 
$\theta_P$ between the $D^0$ momentum and its flight line should be close
to 0, as shown in Fig.~\ref{fig:D0sketch}).  The selection strategy is 
described in detail in Ref.~\cite{D0jpg}.

\begin{figure}[!t]
  \begin{center}
    \includegraphics[clip,width=.49\textwidth]{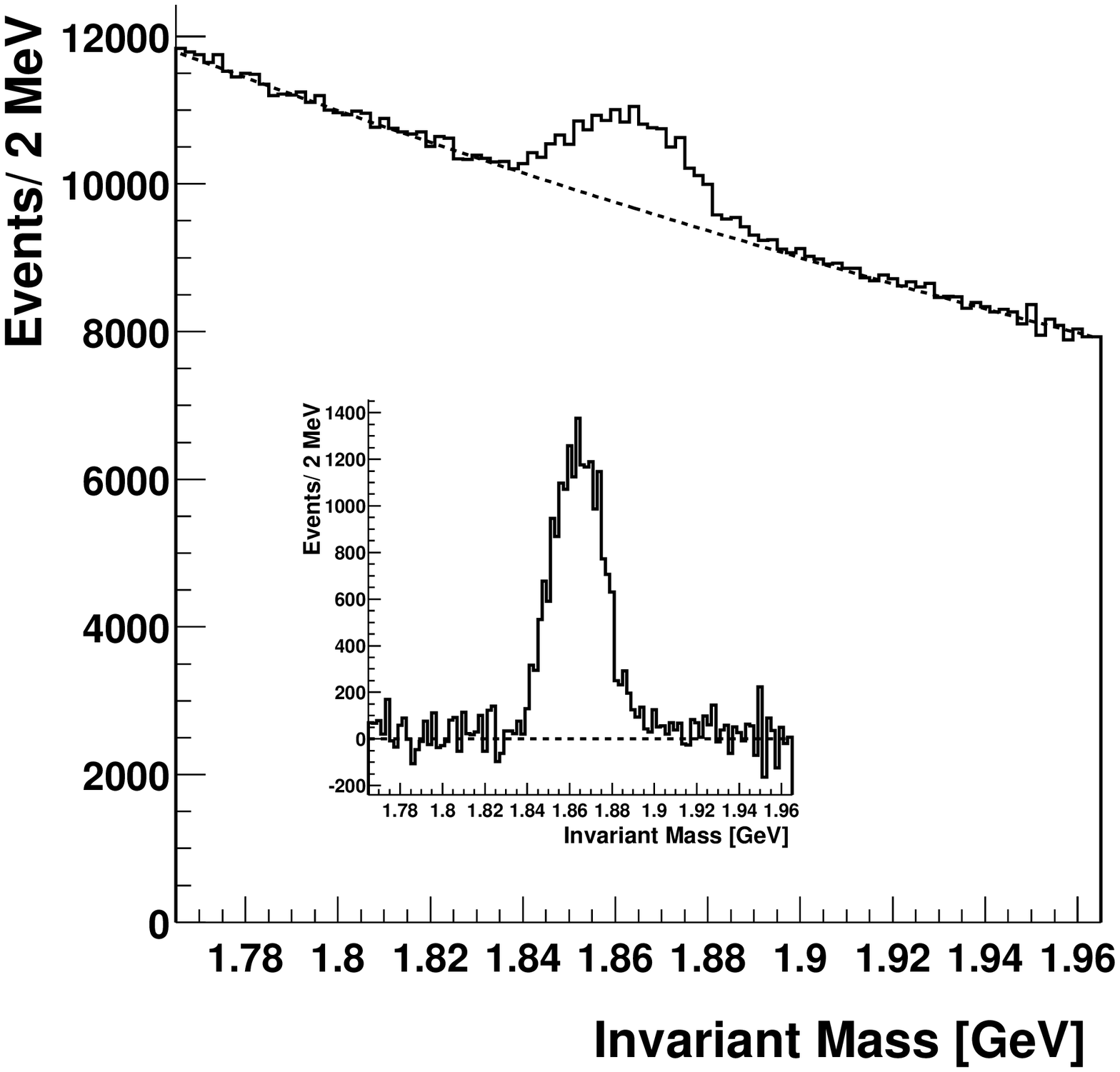}
    \includegraphics[clip,width=.47\textwidth]{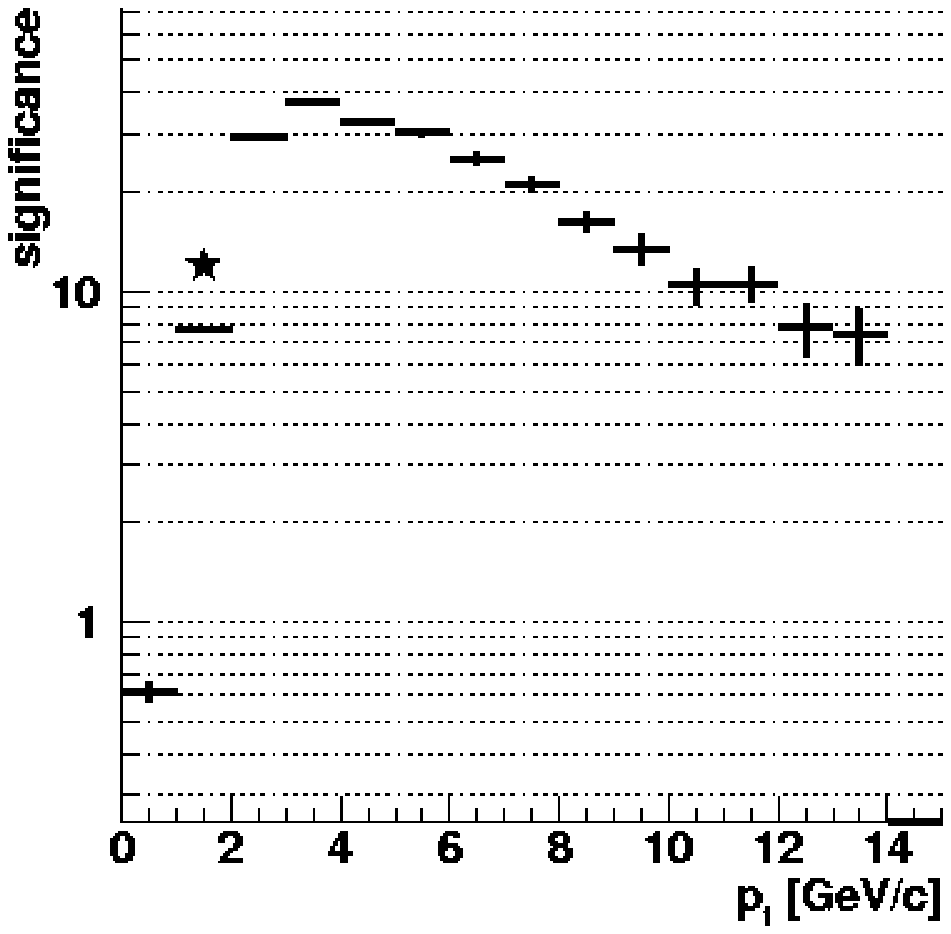}
    \caption{Left: $K\pi$ invariant mass distribution for $10^7$ events 
             (after background subtraction in the inset). 
             Right: statistical significance of the 
             signal as a function of the $D^0$ transverse momentum. 
             The marker shows the significance obtained in the bin 
             $1<p_T<2\ \GeV/c$ requiring the identification of the kaon 
             from its time of flight.} 
    \label{fig:results}
  \end{center}
\end{figure}

After selection cuts, which have been optimized as a function of 
the $D^0$ meson transverse momentum ($p_T$), the ratio $S/B$ is $0.1$ and 
the statistical significance of the signal is $S/\sqrt{S+B}=37$ for 
$10^7$ Pb--Pb events, corresponding to 1 month of data-taking of ALICE.
Figure~\ref{fig:results} (left) shows the corresponding $K\pi$ invariant 
mass distribution. On the same figure (right) the significance is displayed 
as a function of the $D^0$ transverse momentum: the significance is larger 
than 10 up to $p_T\simeq 10\ \GeV/c$. The lower $p_T$ limit of $1\ \GeV/c$ 
will allow a rather safe extrapolation to $p_T=0$ and hence the measurement 
of the total charm production cross section with good accuracy.

\end{document}